\documentclass{appolb}
\usepackage{epsfig}

\begin{document}
\title{Inhomogeneous phases of strongly interacting matter %
\thanks{Presented at the EMMI workshop and
XXVI Max Born Symposium ``Three Days of Strong Interactions'',
Wroc\l aw, Poland, July 9-11, 2009.}%
}
\author{Michael Buballa
\address{Institut f\"ur Kernphysik, Technische Universit\"at Darmstadt,
Germany}
\and
Dominik Nickel
\address{
Institute for Nuclear Theory, University of Washington, Seattle, USA
}
}
\maketitle
\begin{abstract}
We discuss possible inhomogeneous phases 
in two regions of the QCD phase diagram: We begin with color superconducting 
quark matter at moderately high densities, which is an imbalanced Fermi system 
due to the finite strange quark mass and neutrality constraints. Within an 
NJL-type toy model we find that this situation could lead to the formation 
of a soliton lattice.
Similar solutions also exist in the context of the chiral phase transition. 
As an interesting result, the first-order transition line in the phase
diagram of homogeneous phases gets replaced by an inhomogeneous phase
which is bordered by two second-order transition lines.
\end{abstract}
\PACS{12.38.Mh}

\section{Introduction}

The phase structure of QCD is one of the most fascinating
problems in the field of strong interaction physics.
So far, most studies have been performed under the assumption 
that the condensates which define the different phases are homogeneous.
On the other hand, there are good arguments to believe that there
could be inhomogeneous phases as well. 
Well known examples from the literature are,
for instance, the chiral density wave~\cite{Broniowski:1990dy,Nakano:2004cd}, 
the Skyrme crystal~\cite{Goldhaber:1987pb},
and crystalline color superconductors~\cite{Alford:2000ze,
Bowers:2002xr,Casalbuoni:2005zp,Mannarelli:2006fy,Rajagopal:2006ig}.
They also emerge in the large $N$-limit of the $1+1$ dimensional Gross-Neveu 
model \cite{Schnetz:2004vr,Schnetz:2005ih}.

In the main part of this talk we concentrate on color superconductors.
In BCS theory, pairing occurs among fermions with opposite momenta,
forming Cooper pairs with zero total momentum.
If both fermions are at their respective Fermi surface,
the pair can be created at no free-energy cost and the pairing
is always favored as soon as there is an attractive interaction.
This is, however, no longer the case if the Fermi momenta of the 
fermions to be paired are unequal.
This situation arises naturally in quark matter as a consequence 
of the heavier strange quark mass and requirements of neutrality
and beta equilibrium \cite{Alford:2002kj,Steiner:2002gx}.
BCS pairing then requires that the Fermi spheres first have to be 
equalized.
This could be realized, e.g., in a weak 
process which replaces some of the down quarks by strange quarks.
Of course, it will only be favorable if the free energy which is
needed for this process is overcompensated by the pairing energy.
Consequently there is a limit for this mechanism in terms of the 
Fermi momentum difference in the unpaired system and the BCS 
gap~\cite{ClCh1962}.

An alternative option is then that the matter becomes 
inhomogeneous~\cite{Fulde:1964zz,LO64,Casalbuoni:2003wh}.
The basic idea is to form Cooper pairs with non-zero total momentum.
This has the obvious advantage that the fermions in the pair no longer
have to have opposite momenta, and therefore each of them can stay
on its respective Fermi surface.
In the context of color superconductors, this possibility has been
investigated first in Ref.~\cite{Alford:2000ze}.  
The authors restricted themselves to a plane-wave ansatz for 
the gap function, as suggested first by Fulde and Ferrell (FF) in the
context of metallic superconductors~\cite{Fulde:1964zz}. 

On the other hand, since in the FF ansatz the total momentum of the 
pair is restricted to a non-zero but constant value, 
this pairing pattern is strongly disfavored by phase space in most 
cases.
One should therefore consider a multiple plane wave ansatz,
as originally suggested by Larkin and Ovchinnikov (LO) \cite{LO64}.
In the context of color superconductors, this was discussed, e.g., in
Refs.~\cite{Bowers:2002xr,Casalbuoni:2005zp,Mannarelli:2006fy,
Rajagopal:2006ig}.
As expected, the resulting solutions were found to be strongly favored
against the FF phase. 
However, these analyses were restricted to a Ginzburg-Landau 
approximation.
Moreover, the crystal structures have been restricted
to superpositions of a finite number of plane waves whose wave vectors
all have the same length, whereas one should also allow for the
superposition of different wave lengths.
A first step to overcome these shortcomings was done in 
Ref.~\cite{Nickel:2008ng} where inhomogeous pairing was studied
within an NJL model. The main ideas of this paper will be discussed in 
Sec.~\ref{sec:csc}. 

More recently, the idea of having inhomogeneous phases in the 
context of the {\it chiral} phase transition was revisited in 
Refs.~\cite{Nickel:2009ke,Nickel:2009wj} for the NJL model.
In this case the transition from a dynamically broken to the chirally 
restored phase when increasing the chemical potential is delayed by the
formation of an inhomogeneous chirally broken ground state. 
Also the orders of the phase transitions are modified.
This will be discussed briefly in Sec.~\ref{sec:chiral}.

\section{Inhomogeneous color superconductivity}
\label{sec:csc}

\subsection{Model}
 
We consider an NJL-type
Lagrangian for massless quarks $q$ with three flavor 
and three color degrees of freedom,
\begin{equation}
    \mathcal{L} = \bar q \,(i{\partial\hspace{-2.0mm}/} 
    + \hat\mu\gamma^0)\,q + 
    \mathcal{L}_\mathit{int}\,,
\end{equation}
where $\hat\mu$ is the diagonal matrix of chemical potentials. 
The interaction term is given by
\begin{equation}
    \mathcal{L}_\mathit{int} 
    = H \hspace{-3mm} \sum_{A,A'=2,5,7} \hspace{-3mm}
      ( \bar q \,i \gamma_5 \tau_A \lambda_{A'}\, q_C ) 
      ( \bar q_C \,i \gamma_5 \tau_A \lambda_{A'}\,q  )\,. 
\label{Lint}
\end{equation}
Here $H$ is a dimensionful coupling constant and 
$q_C(x) = C\bar q^T(x)$, where
$C=i \gamma^2 \gamma^0$ is the matrix of charge conjugation. 
$\tau_A$ and $\lambda_{A'}$ denote the antisymmetric Gell-Mann matrices 
acting in flavor space and color space, respectively. 
Thus, $\mathcal{L}_\mathit{int}$ corresponds to a quark-quark interaction
in the scalar flavor-antitriplet color antitriplet channel.
Applying standard bosonization techniques, the interaction term,
can equivalently be rewritten as
\begin{equation}
    \mathcal{L}_\mathit{int} 
    = \frac{1}{2}
    \sum_{A,A'} \Big\{\;
     (\bar{q}\, \gamma_5 \tau_A \lambda_{A'}\,q_C)\;\varphi_{AA'} 
     \;+\; h.c. 
      \;-\, \frac{ 1}{2H}\, \varphi_{AA'}^\dagger\,\varphi_{AA'} \Big\},
\label{Lintbos}
\end{equation}
with the auxiliary complex boson fields $\varphi_{AA'}(x)$, 
which, by the equations of motion,
$\varphi_{AA'}(x) = -2H\,\bar q_C(x)\,\gamma_5 \tau_A \lambda_{A'}\,q(x)$,
can be identified with scalar diquarks.
In mean field approximation we replace these quantum fields by their
expectation values
\begin{equation}
    \langle{\,\varphi_{AA'}(x)}\rangle = \Delta_A(x)\,\delta_{AA'}\,,
\end{equation}
where the ``gap function'' $\Delta_A(x)$ is now a classical field. 
Here we assume that the condensation takes place only in the 
diagonal flavor-color components of the gap matrix, $A=A'$, 
as in the standard ansatz for the CFL or the 2SC phase.
Note, however, that we retain the full space-time dependence of the
field. 
Introducing Nambu-Gor'kov bispinors,
$\Psi(x) = \frac{1}{\sqrt{2}} \left(q(x),q_C(x)\right)^T$
and the notation
$\hat\Delta(x) = \sum_{A} \Delta_{A}(x) \,\tau_{A}\lambda_{A}$,
we obtain the effective mean-field Lagrangian
\begin{equation}
    \mathcal{L}_\mathit{MF}(x) = \bar\Psi(x)\,S^{-1}(x)\,\Psi(x) 
    \,-\, \frac{ 1}{4H} \sum_A |\Delta_A(x)|^2\,,
\label{LMF}
\end{equation}
with the inverse dressed quark propagator
\begin{equation}
     S^{-1}(x) = 
\left(
\begin{array}{cc}
i{\partial\hspace{-2.0mm}/}+\hat\mu\gamma^0 & \hat\Delta(x)\, \gamma_{5}\\
-\hat\Delta^{*}(x)\, \gamma_{5} & i{\partial\hspace{-2.0mm}/}-\hat\mu\gamma^0
\end{array}
\right)\,.
\end{equation}
Since $\mathcal{L}_\mathit{MF}$ is bilinear in the Nambu-Gor'kov fields,
they can formally be integrated out and we obtain the mean-field
thermodynamic potential per volume $V$
\begin{equation}
    \Omega_{MF}(T,\hat\mu) 
    \,=\,
    -\frac{1}{2}\frac{T}{V} \mathrm{Tr}\ln\,\frac{S^{-1}}{T}
    \,+\, 
    \frac{T}{V}\sum\limits_A \int\limits_{[0,\frac{1}{T}]\otimes V} 
    \hspace{-3mm}d^4x\;  
    \frac{|\Delta_{A}(x)|^2}{4H}\,.
\end{equation}
Note, however, that the evaluation of the functional $\mathrm{Tr}\ln$
is highly nontrivial because of the $x$ dependent gap functions.

To proceed, we assume that the gap matrix $\hat\Delta(x)$ is time 
independent and periodic in space,
$\hat\Delta(x) \equiv \hat\Delta(\vec x) = \hat\Delta(\vec x + \vec a_i)$,
$i = 1,2,3$.
Hence, $\hat\Delta$ can be decomposed into a discrete set of Fourier 
components
\begin{equation}
    \hat\Delta(x) = \sum_{q_k} \hat\Delta_{q_k}\,e^{-iq_k\cdot x}\,,
\label{Deltaq}
\end{equation}
where the allowed momenta $q_k = (0, \vec q_k)$
form a reciprocal lattice in momentum space.
The inverse propagator is then a matrix whose momentum components
are given by
\begin{equation}
S^{-1}_{p_m,p_n} = 
\left(
\begin{array}{cc}
({p\hspace{-1.7mm}/}_n+\hat\mu\gamma^0)\,\delta_{p_m,p_n} & 
\sum_{q_k}\hat\Delta_{q_k}\gamma_{5}\,\delta_{q_k,p_m-p_n}\\
-\sum_{q_k}\hat\Delta^*_{q_k}\gamma_{5}\,\delta_{q_k,p_n-p_m}& 
({p\hspace{-1.7mm}/}_n-\hat\mu\gamma^0)\,\delta_{p_m,p_n}
\end{array}
\right)\,.
\label{Sinv}
\end{equation}
Note that in general $S^{-1}$ is not diagonal in momentum space
because the condensates $\hat\Delta_{q_k}$ couple different momenta.
Physically, this corresponds to processes like the absorption of 
a hole with momentum $p_n$ by the condensate together with the 
emission of a quark with momentum $p_m = p_n + q_k$.
This is only possible because the inhomogeneous diquark condensates
carry momentum. In the homogeneous case, $\hat\Delta(x) = \mathit{const.}$,
only the momentum component $q_k = 0$ exists, and the in- and outgoing
quark momenta are equal. 
While this is no longer true for our inhomogeneous ansatz, the fact 
that we consider a static solution still guarantees that the {\it energy}
of the quark is conserved. This means, $S^{-1}$ is still 
diagonal in the Matsubara frequencies $\omega_{p_n}$, which can 
therefore be summed in the usual way. 
To that end we write
\begin{equation}
S^{-1}_{p_m,p_n} = \gamma^0\,(i\omega_{p_n} - 
    {\cal H}_{\vec p_m,\vec p_n})\, \delta_{\omega_{p_m},\omega_{p_n}},
\end{equation}
defining the effective Hamilton operator ${\cal H}$,
which does not depend on $\omega_{p_n}$. 
Since ${\cal H}$ is hermitian, it can in principle be diagonalized.
In this context it is important that, since the momenta $\vec q_k$ of
the condensates form a reciprocal lattice, not all momentum components
are coupled with each other, but ${\cal H}$ is block diagonal
with one block ${\cal H}(\vec k)$ for each vector $\vec k$ in the 
Brioullin zone ($B.Z.$).
We then finally obtain for the thermodynamic potential
\begin{equation}
    \Omega_{MF}
    =
    -\frac{1}{4}\hspace{-1mm}
    \int\limits_{B.Z.}\hspace{-1mm} \frac{d^3k}{(2\pi)^3}
    \sum\limits_{\lambda}
    \left[ E_\lambda(\vec k) 
    + 2T\,\ln \left(1 + 2e^{-E_\lambda(\vec k)/T} \right) \right]
    \,+ \sum\limits_A\sum\limits_{q_k} \frac{|\Delta_{A,q_k}|^2}{4H},
\label{Omcont}
\end{equation}
where $E_\lambda(\vec k)$ are the eigenvalues of ${\cal H}(\vec k)$.
Here one should note that ${\cal H}(\vec k)$ is not only an infinite
matrix in momentum space, but each momentum component is also
a $72 \times 72$ matrix corresponding to 4 Dirac, 3 color, 3 flavor, and
2 Nambu-Gor'kov components. 
The numerical studies below are therefore performed in a simplified model,
where the main focus is on the new features related to the inhomogeneity.
To that end we consider a 2SC-like pairing scheme, where only two flavors
(``up'' and ``down'') and two colors (``red'' and ``green'') are paired,
so that the remaining ones (``strange'' and ``blue'') decouple.
We assume that the chemical potentials may be different for up and
down quarks, 
\begin{equation}
    \mu_u = \bar\mu + \delta\mu~, \quad \mu_d = \bar\mu - \delta\mu~,
\label{deltamu}
\end{equation}
but do not depend on color.
Furthermore, we simplify the Dirac structure via a high-density 
approximation (see Ref.~\cite{Nickel:2008ng} for details). 
The problem is then reduced to diagonalize an
effective Hamiltonian whose momentum components are only
$2 \times 2$ matrices,
\begin{equation}
({\cal H}_{\Delta,\delta\mu})_{\vec p_m,\vec p_n} = 
\left(
\begin{array}{cc}
(p_m-\bar\mu-\delta\mu)\,\delta_{\vec p_m,\vec p_n} &  \Delta_{p_m-p_n}
\\
\Delta^*_{p_n-p_m} & -(p_m-\bar\mu+\delta\mu)\,\delta_{\vec p_m,\vec p_n}
\end{array}
\right)~.
\label{HDm}
\end{equation}

Finally, we should note that the thermodynamic potential as defined
in Eq.~(\ref{Omcont}) is divergent and needs to be regularized. 
As discussed in Ref.~\cite{Nickel:2008ng}, a restriction of the momenta
of the effective Hamiltonian, which would be a straight forward 
generalization of the standard momentum cutoff in homogeneous phases, 
leads to strong regularization artifacts. We therefore suggest a
Pauli-Villars like regularization scheme, where the regulator terms are
given by replacing the free-energy eigenvalues $E_\lambda$ by
$E_{\lambda,j} = \sqrt{E_\lambda^2 + j \Lambda^2}$.

\subsection{Numerical results for one-dimensional periodic structures}
 
In the following, we restrict ourselves to $T=0$ and to a fixed 
average chemical potential $\bar\mu = 400$~MeV, so that
$\delta\mu$ is the only remaining external variable.
Our model has two parameters, namely the coupling constant $H$ and the 
cutoff parameter $\Lambda$.
Having fixed the cutoff, we can express the coupling constant 
$H$ through the corresponding value of the BCS gap at $\delta\mu=0$. 
The examples shown below correspond to $\Lambda = 400$~MeV and 
$\Delta_{BCS} = 80$~MeV.
We remind that $\Lambda$ restricts the free energies and not the momenta. 
Thus, the most relevant excitations around the Fermi surface are always
included, and there is no need for $\Lambda$ to be larger than the
chemical potential. 

Our goal is to find the most stable solution, i.e., the minimum
of $\Omega$ with respect to the gap function $\Delta(\vec x)$.
Since the general solution of this problem is rather difficult,
we restrict ourselves to one dimensional modulations, i.e., to
gap functions which vary periodically
in one spatial direction ($z$-direction), but stay constant in the two 
spatial directions ($x$ and $y$),
\begin{equation}
\Delta(z) = \sum_{k} \Delta_{q,k} e^{2ikqz} \,. 
\label{1Dansatz}
\end{equation}
Moreover, we assume that $\Delta(z)$ is real, 
$\Delta_{q,k}^* = \Delta_{q,-k}$.

\begin{figure}[tb]
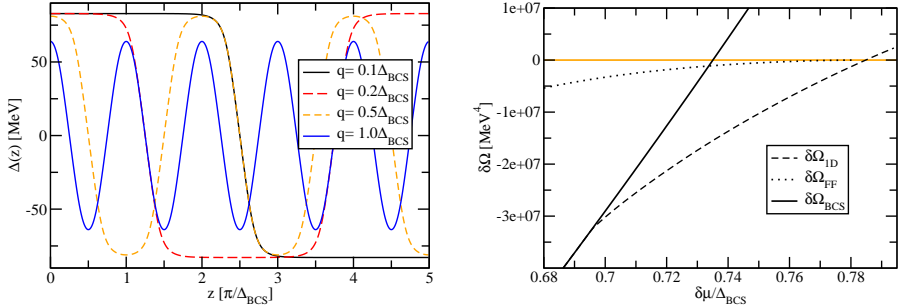

\begin{center}
\epsfig{file=Deltax70v2.eps,width=0.45\textwidth}
\quad
\epsfig{file=Omegadmuv2.eps,width=0.45\textwidth}
\caption{Left: The gap function in coordinate space 
  at $\delta\mu=0.7\Delta_{BCS}$ for different fixed
  values of $q$. 
  Right: Difference between the thermodynamic potentials of 
  different solutions and the normal phase as functions of 
  $\delta\mu$: BCS phase (solid line), general one-dimensional 
  ansatz (dashed line), and single plane wave (FF) 
  ansatz (dotted line). From Ref.~\cite{Nickel:2008ng}.
\label{fig:Deltaq}
}
\end{center}
\end{figure}

In a first step, we take a fixed period, i.e., a fixed value of $q$
and minimize the thermodynamic potential with respect to the 
Fourier components $\Delta_{q,k}$. 
In the left panel of Fig.~\ref{fig:Deltaq} we present examples we 
obtained for $\delta\mu=0.7\Delta_\mathit{BCS}$.
At $q \sim \Delta_{BCS}$ the gap function appears to be sinusoidal. 
For larger periods, however, a new feature becomes apparent: 
the formation of a soliton lattice.
Especially for $q=0.1\,\Delta_\mathit{BCS}$, we see that the gap function 
stays nearly constant at $\pm\Delta_{BCS}$ for about one half-period 
and then changes its sign in a relatively small interval.
The $q=0.2\,\Delta_\mathit{BCS}$ solution behaves qualitatively similar, 
but has a shorter plateau. Remarkably, the shape of the two functions 
is almost identical in the transition region where the gap functions 
change sign. This remains even true for the $q=0.5\,\Delta_\mathit{BCS}$ 
solution, which is kind of an extreme case with no plateau and only 
transition regions. We may thus interprete these transition regions
as very weakly interacting solitons, which are almost unaffected by the
presence of the neighboring (anti-) solitons as long as they do not
overlap. 

It turns out that the gap functions can be fitted remarkably well by
Jacobi elliptic functions, 
$\Delta_{\mathrm{fit}}(z) = A\,\mathrm{sn}\big(\kappa(z-z_0);\nu\big)$,
which is the known shape of the gap functions in $1+1$ 
dimensions~\cite{Machida:1984zz} (see Eq.~(\ref{MoD}) below).
An important difference is, however, 
that in $3+1$ dimensions the amplitude $A$ is not directly related to
the elliptic modulus $\nu$, but must be fitted independently. 

\begin{figure}[tb]
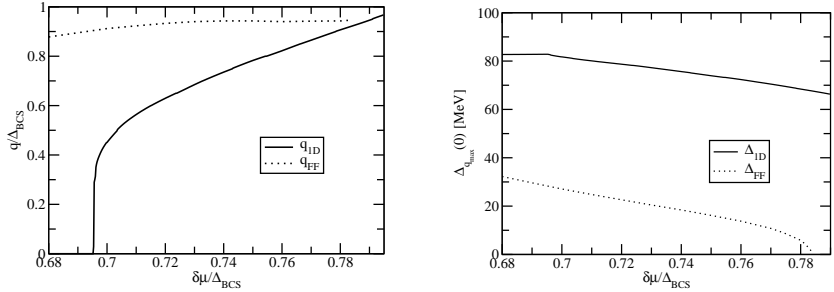

\begin{center}
\epsfig{file=qdmuv4.eps,width=0.4\textwidth}
\qquad
\epsfig{file=Ddmuv2.eps,width=0.4\textwidth}
\caption{The energetically preferred value of $q$ (left) 
and the corresponding amplitude of the gap function (right)
as functions of $\delta\mu$. Solid line: general one-dimensional
ansatz, dotted line: single plane wave (FF) ansatz. From 
Ref.~\cite{Nickel:2008ng}.
\label{fig:qamp}
}
\end{center}
\end{figure}

\begin{figure}[tb]
\begin{center}
\epsfig{file=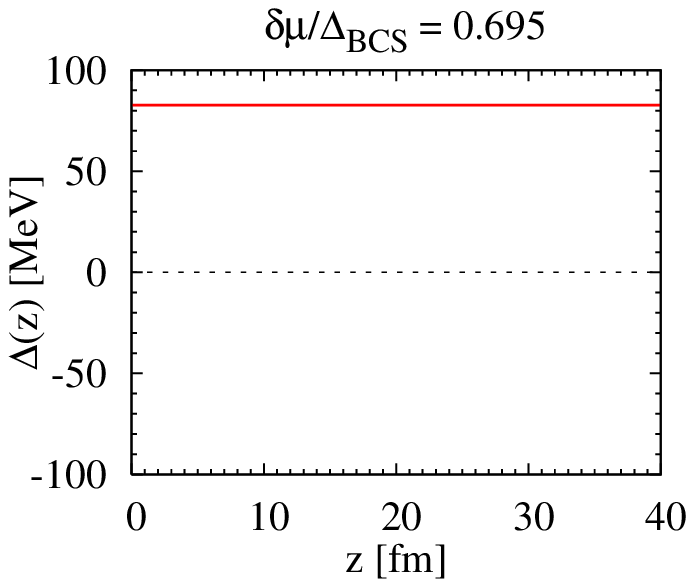,width=0.24\textwidth}
\epsfig{file=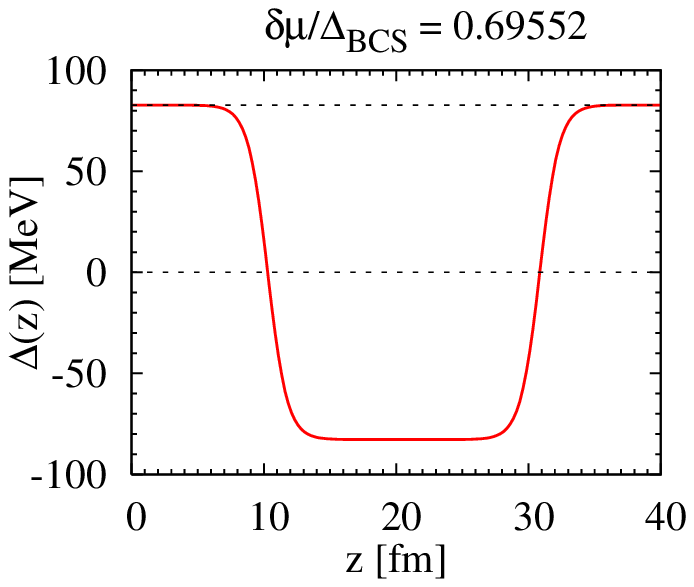,width=0.24\textwidth}
\epsfig{file=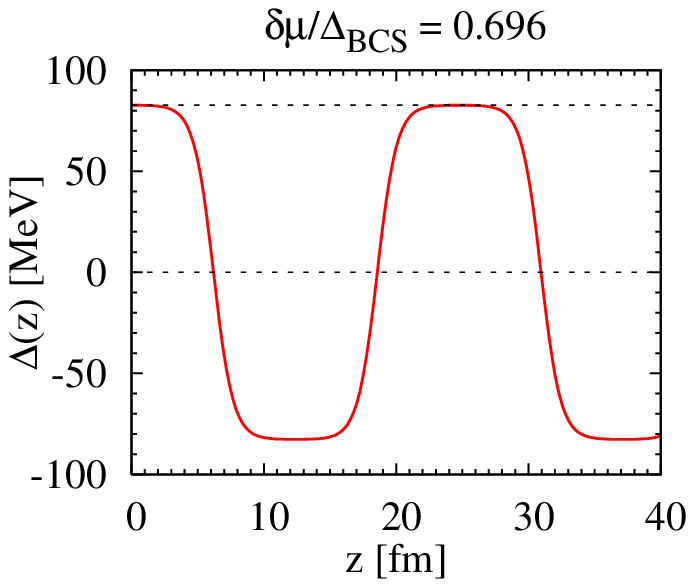,width=0.24\textwidth}
\epsfig{file=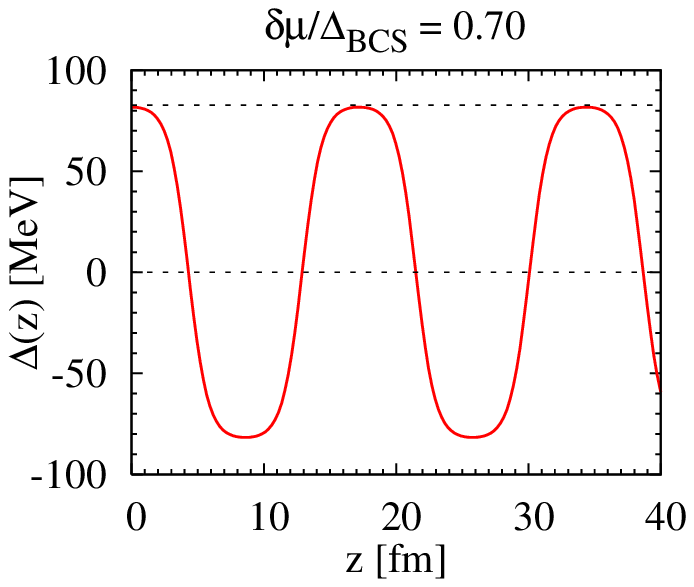,width=0.24\textwidth}
\epsfig{file=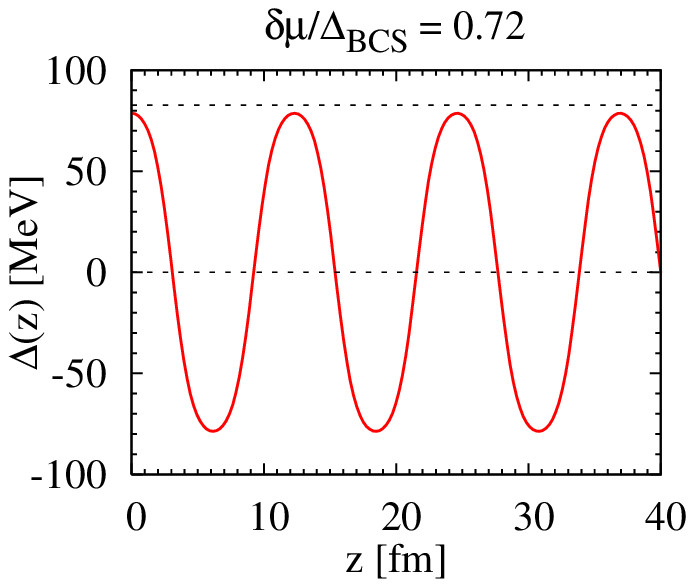,width=0.24\textwidth}
\epsfig{file=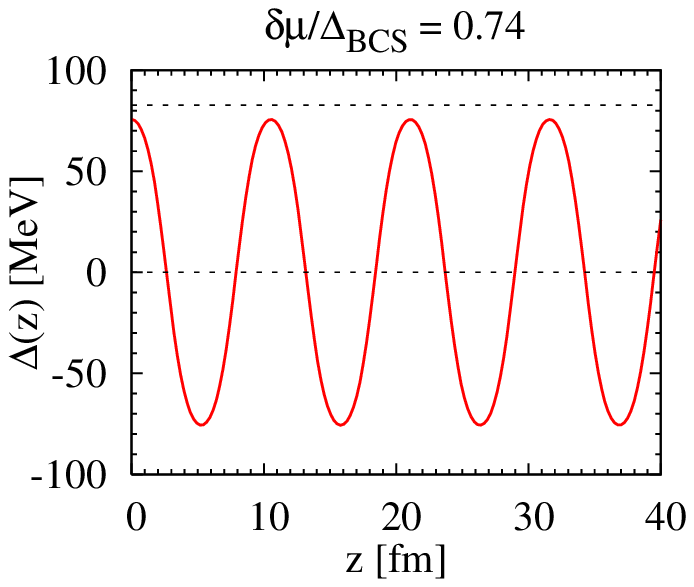,width=0.24\textwidth}
\epsfig{file=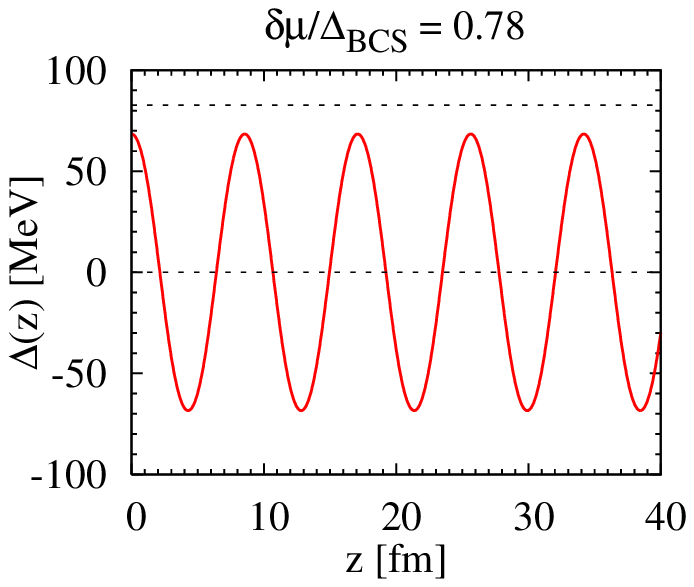,width=0.24\textwidth}
\epsfig{file=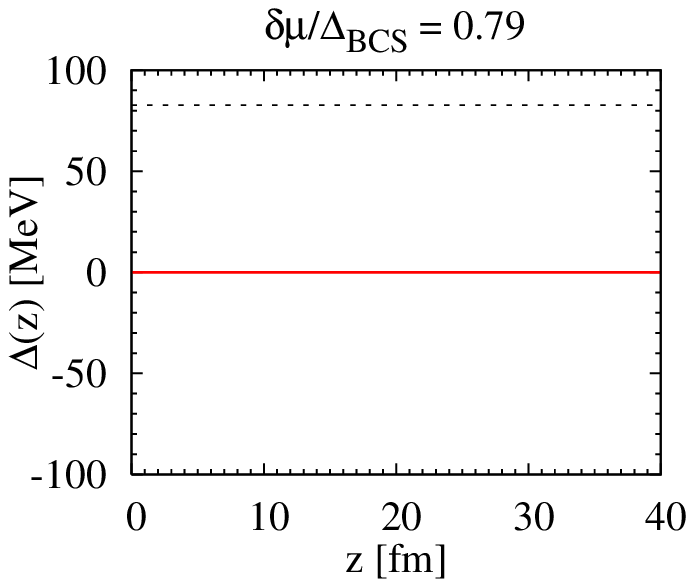,width=0.24\textwidth}
\caption{The color superconducting gap function in coordinate space for 
         different values of $\delta\mu$ and the corresponding most favored 
         value of $q$. 
\label{fig:Deltax}
}
\end{center}
\end{figure}

For each $\delta\mu$, 
with the solutions for the different chosen values of $q$ at hand,  
we now have to minimize the thermodynamic potential in $q$.
The resulting free energy gain compared to the normal conducting solution
is shown in the right panel of Fig.~\ref{fig:Deltaq}.
While the BCS solution (solid line) and the normal solution
are favored at low and high $\delta\mu$, respectively, there is a
window at intermediate $\delta\mu$ where the inhomogeneous solution is 
favored. Note that for the general one-dimensional ansatz (dashed line),
this window is about twice as wide as it would be for the single 
plane-wave ansatz (FF phase, dotted), which is energetically much less favored.
Most striking, whereas the BCS-FF phase transition would be first order,
the transition from the BCS phase to the general one-dimensional phase 
is second order. This is possible because the most favored value of $q$ 
in the inhomogeneous phase goes continuously to zero, when $\delta\mu$
is reduced towards the phase boundary, see left panel of Fig.~\ref{fig:qamp} 
(solid line). 
The corresponding amplitude of the gap function is displayed on the 
right-hand side of Fig.~\ref{fig:qamp}. Unlike the FF amplitude (dotted),
the amplitude of the general one-dimensional gap function stays
of the order of $\Delta_\mathit{BCS}$ at large $\delta\mu$. As a consequence,
the phase transition to the normal phase is first order.
The resulting behavior of the gap function with increasing $\delta\mu$
is sketched in Fig.~\ref{fig:Deltax}.

\section{Chiral Phase transition}
\label{sec:chiral}

More recently, a similar study of inhomogeneous phases has been performed 
for the chiral phase transition~\cite{Nickel:2009ke,Nickel:2009wj}.
Here, instead of the color superconducting gap, the chiral condensate
or, equivalently, the ``constituent quark mass'' was allowed to have
general one-dimensional periodic modulations.
This case actually turned out to be simpler, because it can be shown
that the analytically known mass functions 
\begin{equation}
 M_{1+1}(z) = \sqrt{\nu}\kappa\,\mathrm{sn}\big(\kappa(z-z_0);\nu\big)
\label{MoD}
\end{equation}
of the $1+1$ dimensional Gross-Neveu model~\cite{Schnetz:2004vr}
can be lifted to $3+1$ dimensions by Lorentz boosts~\cite{Nickel:2009wj}.
Apart from the irrelevant shift $z_0$, this function depends only
on two parameters $\kappa$ and $\nu$, which are related to the wave
vector $q = \pi\kappa/4K(\nu)$ ($K(\nu) =$ complete elliptic integral
of the first kind) and the amplitude $\sqrt{\nu}\kappa$.
Thus, at each temperature and chemical potential, the thermodynamic 
potential must be minimized w.r.t. $\kappa$ and $\nu$.\footnote{ 
Eq.~(\ref{MoD}) corresponds to the chiral limit. It is possible to
to generalize this solution to finite current quark 
masses~\cite{Schnetz:2005ih}, introducing a third parameter.}

\begin{figure}[tb]
\begin{center}
\epsfig{file=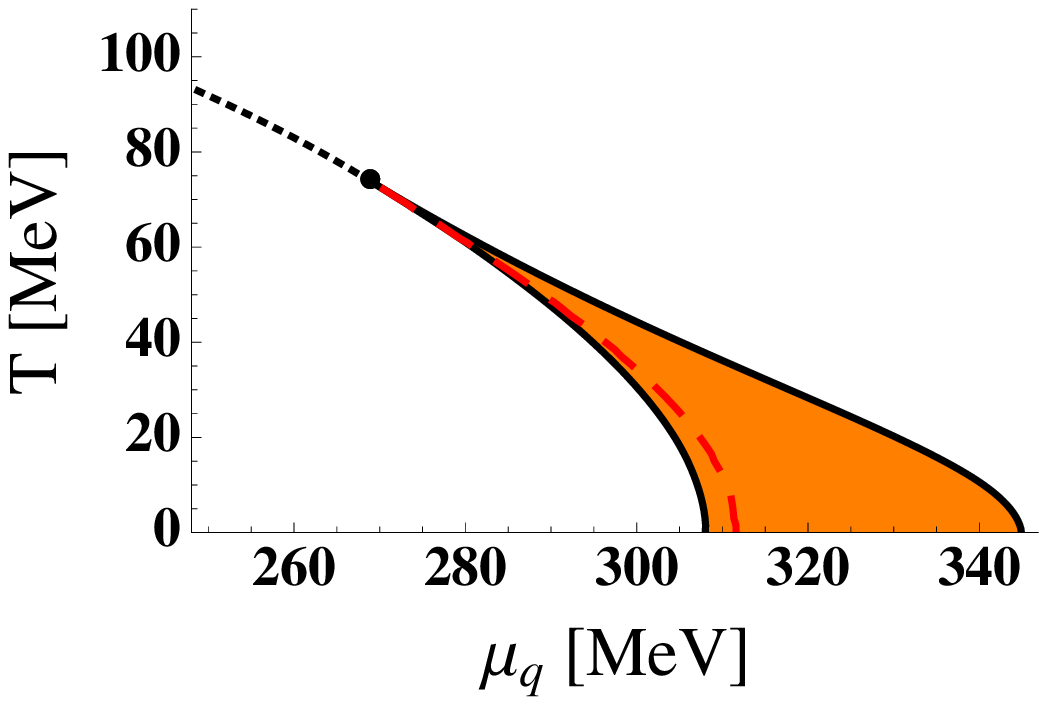,width=0.4\textwidth}
\qquad
\epsfig{file=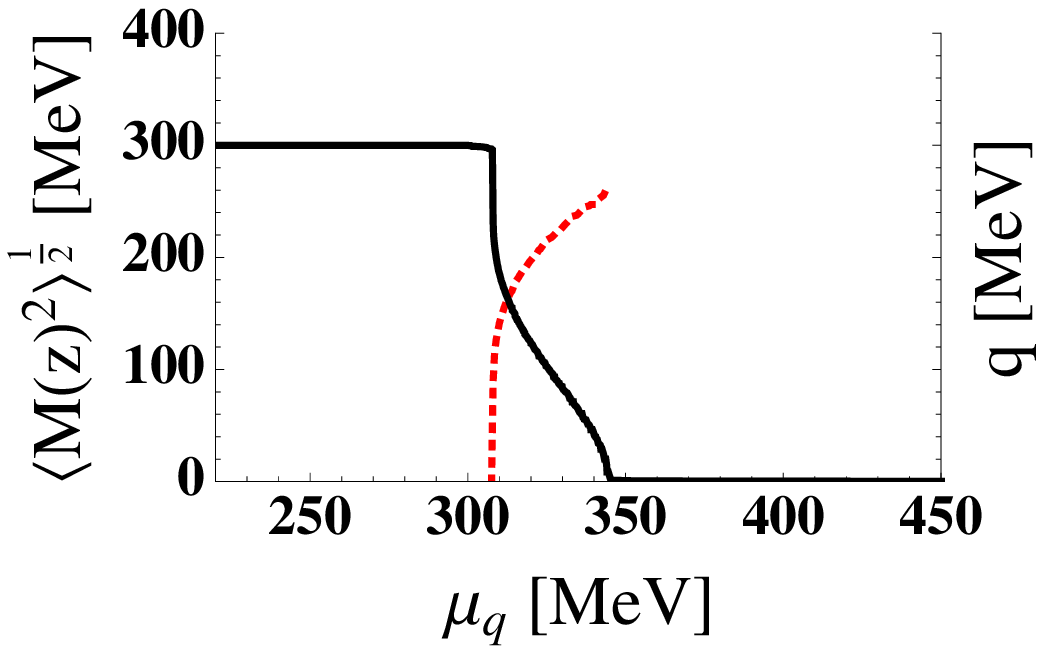,width=0.4\textwidth}
\caption{Left: Phase diagram in the chiral limit. The orange shaded 
region marks the inhomogeneous regime. 
Right: Wave vector $\vert\vec{q}\vert$ (dashed line) and average amplitude 
$\sqrt{\langle M^2\rangle}$ (solid line) at $T=0$ as functions of the
quark chemical potential $\mu_q$. Adapted from Ref.~\cite{Nickel:2009wj}.
\label{fig:chiral}
}
\end{center}
\end{figure}

The resulting phase diagram for a two-flavor NJL model 
in the chiral limit is shown in the left panel 
of Fig.~\ref{fig:chiral}. When the solutions are restricted to homogeneous
condensates, the phase transition is first order at lower temperatures
(red long-dashed line) and second order at higher temperatures (black dotted
line). However, when inhomogeneous condensates are taken into account, 
the first-order phase boundary becomes completely hidden by an inhomogeneous
phase (orange shaded region).
The latter is bordered by two second-order boundaries, whose intersection
defines a Lifshitz point. In the NJL model the Lifshitz point precisely agrees
with the critical point~\cite{Nickel:2009ke}, 
similar to the findings in the Gross-Neveu model.

On the r.h.s.\@ of Fig.~\ref{fig:chiral}, we display the wave vector
$\vert\vec{q}\vert$ (dashed line) and the average amplitude of the mass
function (solid line) as functions
of the chemical potential $\mu_q$ at $T=0$. 
Similar to the superconducting case, $\vert\vec{q}\vert$ goes to zero at 
the lower end
of the inhomogeneous phase, so that the latter is continuously connected 
to the homogeneous phase with broken chiral symmetry. 
At the upper end, the amplitude goes to zero, and 
the inhomogeneous phase is smoothly connected to the chirally restored
phase.

\section{Outlook}

The results presented here should be considered as first steps towards
a more complete picture of the phase diagram with inhomogeneous phases,
and much remains to be done. 
In the context of the chiral phase transition one should couple the quarks
to the Polyakov loop and study the influence of other interaction
channels~\cite{CBN}. On the color-superconductivity side, one should
calculate the equation of state for globally neutral quark matter in
beta equilibrium and extend the analysis to finite temperature. 
Eventually, superconducting and chiral condensates should
be treated simultaneously to obtain a unified phase diagram. 
Moreover, the model should be extended to $2+1$ flavors.

The restriction to inhomogeneities with one-dimensional 
modulations should not be the last word, but it would be interesting to
study two- or even three-dimensional modulations as well. 
Technically, this is of course much more demanding. Eventually, however,
one would like to study the transition from a single nucleon (described
as a single soliton \cite{Christov:1995vm,Alkofer:1994ph})
to nuclear matter and finally to color-superconducting quark matter. 
 
\section*{Acknowledgments}
M.B. thanks the organizers for an inspiring workshop.
D.N. was supported in part by 
the Department of Energy (DOE) under grant numbers DE-FG02-00ER41132 and
DE-FG0205ER41360, furthermore by the German Research Foundation (DFG) under
grant number Ni 1191/1-1.

\end{document}